\documentstyle[]{mn}
\input psfig.sty
\title[Diffuse radio source in Abell 85: estimation of cluster magnetic field]%
{The diffuse, relic radio source in Abell 85:
estimation of cluster scale magnetic field from
inverse Compton X-rays}
\author[Bagchi et al.]%
{J. Bagchi,$^{1}$ V. Pislar,$^{2}$  G.B. Lima Neto$^{2}$\\
$^1$ Indian Institute of Astrophysics, Koramangala, Bangalore 560 012, India\\
$^2$ Institut d'Astrophysique de Paris, CNRS, 98bis Bd Arago,
F-75014 Paris, France\\
}
\date{Accepted ???. Received ????; in original form ????}
\pagerange{\pageref{firstpage}--\pageref{lastpage}}
\pubyear{1998}

\begin{document}
\maketitle
\label{firstpage}
\begin{abstract}

We report the first detection of an inverse Compton X-ray emission,
spatially correlated with a very steep spectrum radio source (VSSRS),
0038-096, without any detected optical counterpart, in cluster Abell~85. 
The ROSAT PSPC data and its multiscale wavelet analysis reveal a large scale 
(linear diameter of the order of 500 $h^{-1}_{50}$ kpc), diffuse 
X-ray component, in excess to the thermal bremsstrahlung, overlapping 
an equally large scale VSSRS. 
The primeval 3 K background photons, scattering off the relativistic electrons 
can produce the X-rays at the detected level. The inverse Compton flux is 
estimated to be $(6.5\pm 0.5)\times 10^{-13}$ erg~s$^{-1}$~cm$^{-2}$ in the 
0.5--2.4~keV X-ray band. A new 327 MHz radio map is presented for the cluster 
field. The synchrotron emission flux is estimated to be $(6.6\pm 0.90) \times 
10^{-14}$ erg s$^{-1}$ cm$^{-2}$ in the 10--100 MHz radio band. The positive 
detection of both radio and X-ray emission from a common ensemble of relativistic 
electrons leads to an estimate of $(0.95\pm 0.10) \times 10^{-6}$ G for the 
cluster-scale magnetic field strength. The estimated field is free of the 
`equipartition' conjecture, the distance, and the emission volume. Further, the 
radiative fluxes and the estimated magnetic field imply the presence of 
`relic' (radiative lifetime $\ga 10^{9}$ yr) relativistic electrons 
with Lorentz factors $\gamma \approx$ 700--1700, that would be a  
significant source of radio emission in the hitherto unexplored frequency range $\nu 
\approx$ 2--10 MHz.
\end{abstract}


\section{Introduction}\label{intro}

Apart from dark-matter, the diffuse intracluster medium has the two main
constituents: the 
bremsstrahlung emitting hot ($T \sim 10^{7-8}$ K), tenuous ($n_0\sim 10^{-(3-4)}$ 
cm$^{-3}$), and diffuse thermal gas, and the higher energy, relativistic
particles (cosmic rays), with 
electrons emitting the magneto-bremsstrahlung (synchrotron) radiation. The 
presence of diffuse, large scale (0.2--1.0 Mpc) synchrotron sources with very 
steep spectra, is known in several clusters (e.g. Coma, Abell 2256, 2319, 85, 
etc.). The origin of these relativistic particles is currently not well understood but
in view of the absence of any optical counterparts, these radio sources 
are believed to be the remnants (`relics' and `halos') of once 
active radio galaxies. These remnants are 
prevented from rapid fading from expansion by the thermal pressure of the surrounding intra 
cluster gas (Baldwin and Scott 1973). The main energy losses of 
their relativistic electrons come from 
synchrotron emission and the inverse Compton scattering of the 3K background 
radiation. Their presence imply that magnetic fields on similar large scales 
may exist in the intracluster space (see Feretti \& Giovannini 1996, and 
Kronberg 1994 for reviews). That this magnetic field on large scales, may be a 
general property of clusters of galaxies, is suggested by the Faraday rotation data on radio 
sources observed in the direction of several clusters (Kim et al. 1991, Kronberg 
1994). Estimates of the field strengths, based on Faraday rotation (Kim et al. 
1991) or the `equipartition' hypothesis (e.g. Feretti \& Giovannini 1996), both 
give a value $B \sim$ 0.5 to 1.0~$\mu$G. Currently, the origin and evolution of these
fields are not well understood due to observational difficulties in estimating the magnetic 
fields with sufficient accuracy and to the small number of
actual estimates that has been attempted so far for a few clusters only (Kronberg 1994). 

Another method of considerable merit to estimate the cluster scale magnetic field
is the detection of co-spatial inverse Compton (IC) X-ray emission with the 
synchrotron emission plasma. The ubiquitous 3K microwave background photons, 
scattering off the relativistic electrons (the IC/3K process), should produce a diffuse 
X-ray `glow' associated with the radio plasma (Feenberg \& Primakoff 1948;
Hoyle 1965; Baylis et al. 1967; 
Felten \& Morrison 1966; Harris \& Grindlay 1979; Rephaeli \& Gruber 1988). The
non-relativistic (thermal) analogue of this process is the well known
Sunyaev-Zel'dovich effect (Rephaeli 1995). A possible detection of
inverse Compton X-ray photons from the scattering of 3K microwave background
would not only provide a magnetic field estimate based on a different physical
process, but would also provide information on as yet unknown non-thermal
X-ray component in galaxy clusters.
The inverse Compton method has
several advantages: both the magnetic field strength, and the electron energy 
spectrum are obtainable from the observed IC/3K and the synchrotron emission 
fluxes, and the field so obtained is independent of the `equipartition' conjecture 
(Pacholkzyck 1970), the distance, and the emission volume. 

Despite a number of attempts over the last 20 years, the detection of 
IC/3K radiation has proved elusive, very often with only upper limits to the X-ray 
fluxes and lower limits to magnetic fields (e.g. Harris \& 
Romanishin 1974, Rephaeli 1977, Rephaeli et al. 1987,
Rephaeli \& Gruber 1988, Bazzano et al. 1990, Rephaeli et al. 1994, Harris et al. 1995).
The principal constraints arise from rarity of diffuse, steep spectrum
cluster radio sources, limited sensitivity, and spatial/spectral 
resolution of low frequency (10--300 MHz) radio or X-ray telescopes, and 
confusion from cluster thermal emission. However, correlating the radio and 
the EINSTEIN satellite data, Bagchi (1992) suggested the possibility of IC/3K 
emission in Abell~85. Recently, using the ROSAT PSPC data,
Laurent-Muehleisen et al. (1994) and Feigelson et al. (1995) claimed the
detection of IC/3K X-ray emission associated with the radio lobes of the
galaxy Fornax-A, further supported
by Kaneda et al. (1995), employing the ASCA X-ray spectral data.

This work employs the good spatial resolution, high sensitivity and spectral 
capabilities of the ROSAT X-ray satellite, to search for co-spatial IC/3K emission 
from the diffuse radio source 0038-096 located in the rich cD cluster Abell~85.
We present evidence for what might be the first detection of the IC/3K emission
from the `relic' relativistic electrons residing in the intra-cluster medium, presumably
the remnants of an once active radio galaxy that is presently unidentified. We 
also present a new 327 MHz radio map of the cluster field. Combining the radio and X-ray 
data, we estimate the magnetic field strength in the co-spatial emission volume. 
We then briefly discuss its implications. For a cluster redshift of 
0.0555 (Pislar et al. 1997), 1 arcmin corresponds to $97 h^{-1}_{50}$ kpc, with  the 
Hubble constant expressed in units of 50~km~s$^{-1}$~Mpc$^{-1}$.

\section{The data}

\subsection{The OSRT Radio Observations}

Abell 85 was observed with the Ooty Synthesis Radio Telescope (OSRT; Sukumar et 
al. 1988), at a frequency of 326.5 MHz (band width 4 MHz), three times during 1985--86. After 
initial calibration, the visibility data were combined and `self-calibration' (Schwab 1980) 
performed to further improve the quality. The calibrated data  was Fourier transformed and 
the image was deconvolved 
with `CLEAN' algorithm (using AIPS package), applied to the entire OSRT field of view
($\sim$ 3 deg. EW x 0.75 deg. 
NS). The final image was convolved with a `clean' beam of 60 arcsec circular 
Gaussian (FWHM) profile. The 1-$\sigma$ dispersion of noise on this image was 
6.5 mJy/beam area. This radio map is shown in Figs. 1 and 2. The OSRT has
the necessary short baseline sampling  to map the $\sim$5 arcmin
angular extent of the VSSRS.

\subsection{The ROSAT X-ray Observations}

\begin{figure}
\centerline{\psfig{figure=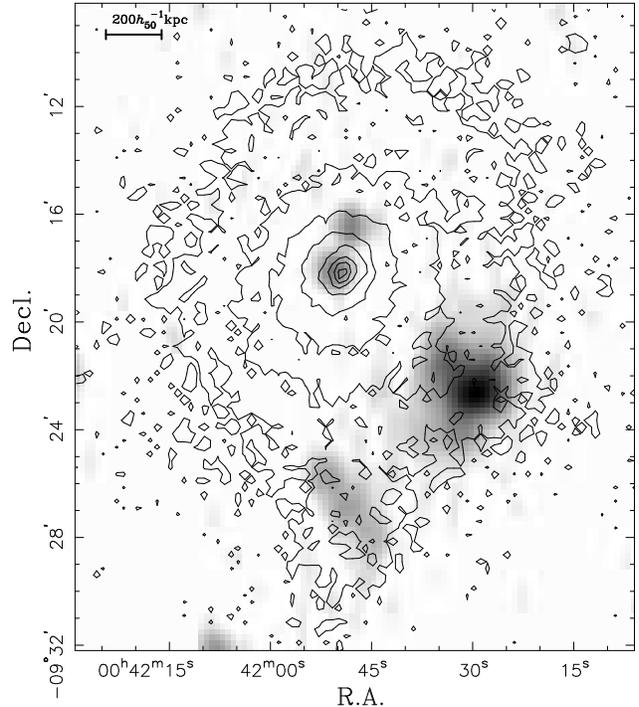,width=8.3truecm}}
\caption{Iso-count contour map for the X-ray emission in Abell 85 (PSPC 0.1--2.4 keV data),
superposed
on OSRT 326.5 MHz radio map (in grey scale). The   
contours are spaced logarithmically. The radio 
source 0038-096 is visible at RA(J2000) $\sim 00^{\rm h}~41^{\rm m}~30^{\rm s}$,
Decl. $\sim -09^\circ~23\arcmin$. Note the excess X-ray emission at this
location. Another correlated X-ray and radio emission could be seen over the
`south-blob' at RA $\sim 00^{\rm h}~41^{\rm m}~48^{\rm s}$, Decl. $\sim
-09^\circ~27\arcmin~30\arcsec$.
}
\protect\label{fig1}
\end{figure}

The Abell~85 field was observed twice (PI Schwarz and PI Jones) by the ROSAT Position
Sensitive Proportional Counter (PSPC). 
We have merged these two images to form one of effective exposure 
of 15949~s using the EXSAS package (Zimmermann et al. 1994). The iso-contours of 
this merged image are shown in Fig.~1. The PSPC has a point spread function (PSF) of
$\approx 25$ arcsec (FWHM) within the central $\approx 10$ arcmin radius and at
the 1~keV energy.

We have processed the merged PSPC image using the multiscale wavelet 
reconstruction technique as described in Pislar et al. 1997 (and references 
therein). With the wavelet analysis, the noise is removed and we can detect 
structures (at different scales) significant at least at the 3 sigma level. The  
wavelet reconstruction of the PSPC data is shown in Fig.~2.

To obtain the X-ray spectrum of Abell~85 (which we will need for the calculation 
of the IC/3K emission, cf. below), we have used the EXSAS package, taking the 
background in a circle of radius 6 arcmin, without any visible sources, 31 
arcmin away from the centre of the cluster. The PSPC image was corrected for 
vignetting and the spectrum rebinned to obtain a signal to noise ratio of 5.

\section{The Radio and X-ray Structures, and the Fluxes}

\begin{figure}
\centerline{\psfig{figure=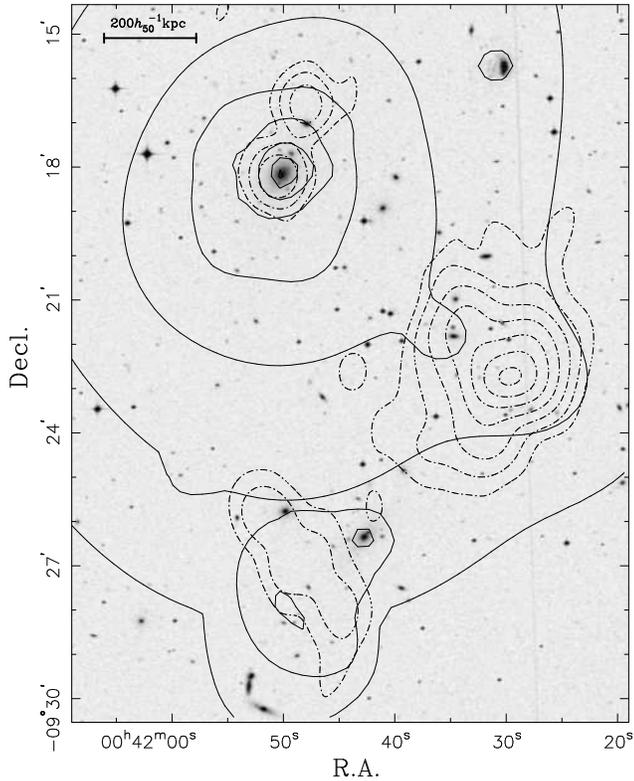,width=8.3truecm}}
\caption{Abell 85 central region at different wavelengths. The photographic
R-filter image (UK Schmidt Telescope and the Digitized Sky Survey) is shown in
grey scale. Full contour lines show the
multiscale wavelet reconstruction of the ROSAT PSPC X-ray data shown in Fig. 1.
The OSRT 326.5 MHz radio surface-brightness is depicted using dot-dashed
contour lines. All contours are spaced logarithmically. 
}
\protect\label{fig2}
\end{figure}

The correlated X-ray emission with the diffuse VSSRS at $\alpha = 00^{\rm 
h}~41^{\rm m}~29\fs5$, $\delta = -09^\circ~22\arcmin~41\arcsec$ (J2000) can be 
seen in Fig.~1 (PSPC data) and Fig.~2 (wavelet reconstruction). The excess 
X-ray component is also diffuse, and has a similar large scale structure (about  
$500 h_{50}^{-1}$ kpc across). The lack of optical identification (Fig.~2), 
and the extremely steep, curved radio spectrum of VSSRS (Fig.~3) suggest that it 
may be a `relic' or `halo' type source of 
undetermined origin, undergoing significant radiative losses
(Slee \& Reynolds 1984, Swarup 1984, Joshi et al. 1986, Bagchi 1992). We note that the 
excess X-ray component can also be well discerned in the ROSAT HRI data
(Lima Neto et al. 1997), and in the EINSTEIN observatory IPC image (Forman \& Jones 1982). 

To estimate the IC/3K X-ray flux of VSSRS, we have carefully subtracted the 
thermal X-ray contribution from its vicinity using the detailed thermal bremsstrahlung 
model given by Pislar et al. (1997). From the model, we evaluated the 
parameters (ellipticity and position angle) of the elliptical iso-intensity 
contour passing over the VSSRS. Then using the actual PSPC counts, the 
bremsstrahlung flux was evaluated, using EXSAS, in 11 circular regions
(excluding the VSSRS, the 
`south blob', and the Seyfert galaxy at $\alpha$ (J2000) $= 00^{\rm 
h}~41^{\rm m}~30\fs4$, $\delta = -09^\circ~15\arcmin~48\arcsec$),
of 5 arcmin diameter each ($\approx$ the VSSRS 
diam.), placed on the iso-intensity contour.
Optical spectroscopic analysis indicates the presence
of a foreground group located towards the west  of Abell~85 (Pislar et
al. 1997); therefore we 
have proceeded in the following way: 7 of the 11 circles are taken on the north 
and east sides of the cluster, and the remaining 4 circles are located over the 
foreground group (where the counts are marginally higher). We have fixed the
neutral hydrogen column density at each point at the 
galactic value (Dickey \& Lockman 1990).

\begin{figure}
\centerline{\psfig{figure=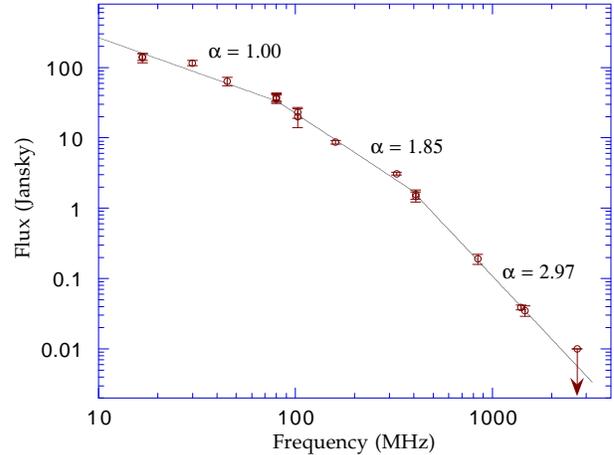,width=8.0truecm}}
\caption{The radio spectrum of VSSRS 0038-096. The piecewise 
least-square spectral fits to the data are
shown by three power laws with the spectral index $\alpha$ indicated alongside. The 326.5 MHz 
data point is the
OSRT measurement ($3.15 \pm 0.15$ Jy). Other data are from published literature.
}
\protect\label{fig3}
\end{figure}

After averaging the 11 measurements with weights appropriate to each (that is, 
depending on the region where it was taken), we obtained: $f_{th}= (1.12 \pm 0.05) 
\times 10^{-12}$ erg~s$^{-1}$~cm$^{-2}$, the thermal bremsstrahlung X-ray flux. We expressed 
the total flux over the VSSRS as the sum of $f_{th}$, and the non-thermal IC/3K 
flux, $f_{IC}$. The $f_{IC}$ was obtained by fixing $f_{th}$ at the above value 
and estimating $f_{IC}$ by a least-square fit to the total flux over the 
VSSRS. A priori, we have assumed that the IC/3K flux has the same spectral index 
$\alpha$ = 1.00 (flux $\propto \nu^{-\alpha}$) as the low frequency 
radio spectral index (Fig. 3), and have put the temperature of the 
bremsstrahlung emitting gas at the mean value of the 11 circular regions ($T = 
2.8$ keV). This results in $f_{IC} = (6.5 \pm 0.5)\times 10^{-13}$ 
erg~s$^{-1}$~cm$^{-2}$ in the 0.5--2.4 keV band. The formal significance of 
$f_{IC}$ detection is about 9 $\sigma$ (random errors).

The radio spectrum of VSSRS 0038-096 is shown in Fig. 3. For the present work, the flux 
density and the spectral index at about 10~MHz are important. This is because 
an electron 
with Lorentz factor $\gamma \approx 1500$ would scatter the average $\nu_{bg} \approx 1.6 
\times 10^{11}$ Hz microwave background photon (near the peak of Planck distribution) 
into the $\sim$ 2 keV ($\nu_{x} 
= (4/3)\;\nu_{bg} \gamma^{2}=4.8 \times 10^{17}$ Hz) X-ray band, and would 
produce the synchrotron emission at $\nu_{syn} \approx$ 10--20 MHz in a magnetic 
field of $ B \sim$ 1--2 $\mu$G ($ \nu_{syn} = 4.19 \ B \gamma^{2}$ Hz). 
Fortunately, the spectrum is known near this frequency range. A 
power-law of spectral index $\alpha = 1.00 \pm 0.10$
gives a good fit to the VSSRS spectrum in the frequency range 10--100 
MHz (Fig.~3). The synchrotron flux, $f_{S} = (6.6 \pm 0.9) \times 10^{-14}$
erg~s$^{-1}$~cm$^{-2}$, is obtained by integrating in the 10--100 MHz frequency 
range.

\section{The Magnetic Field Estimate}

Consider a population of relativistic electrons emitting the synchrotron and 
the IC/3K radiation. Given a 
power-law photon emission spectrum of index $\alpha$ = 1.00 (flux $\propto \nu^{-\alpha}$), 
the energy spectrum 
of electrons is another power law of index $p = 2 \alpha + 1 = 3.00$; d$N(E) = 
N_{0} E^{-p}$d$E$, where $N_{0}$ is the amplitude, and d$N(E)$, the differential 
number density within energy $E$ to $E+$d$E$ (Blumenthal and Gould, 1970). Following the exact 
theoretical derivations by Blumenthal and Gould (1970), the equations for synchrotron flux 
$f_{S}$, and the IC/3K flux $f_{IC}$ can be put in the following practical forms (in cgs 
units, for the indicated bandwidths, and $\alpha = 1.00$ photon sp. index):
\begin{eqnarray}
f_{S} =& 1.64 \times 10^{-14} \frac{N_{0}\; V}{4 \pi D^{2}} B^{2} a(3)&
{\rm (10-100\ MHz)} \nonumber\\
f_{IC} =& 1.36 \times 10^{-29} \frac{N_{0}\; V}{4 \pi D^{2}} T^{4} b(3)&
{\rm (0.5-2.4\ keV)}\, .
\end{eqnarray}
Here, $V$ is the emission volume, $D$, the source distance, $B$, the magnetic field 
strength (randomly oriented in the intra-cluster medium), and 
$T$ = 2.877~K, is the radiation temperature at 
redshift 0.0555 (from COBE measure of T=2.726 K at  z=0; Mather et al. 1994). The factors 
$a(3)$ = 0.0742, and $b(3)$ = 11.54 are derived in 
Blumenthal and Gould (1970). The field B was estimated from these equations by 
substituting the observed values of $f_{S}$ and $f_{IC}$:
\begin{equation}
B = 2.97  \times 10^{-6}  ({f_{S} \over {f_{IC}}})^{1 \over {2}} {\rm G} \,
= (0.95 \pm 0.10) \times 10^{-6} {\rm G}\, .
\label{champs}
\end{equation}

This estimate of $B$ is free of the uncertain quantities $N_{0}$, $V$, and $D$, 
being proportional to the ratio $(f_{S} / f_{IC})^{1 \over 2}$. The `$1 / 2$'
power dependence ensures that the estimated $B$ value would be only weakly 
affected by the errors in the radio and X-ray fluxes.
Note that this field estimate is independent of the  bandwidths
chosen for the X-ray and radio photons, as long as their power law spectral slopes are equal.

\section{Discussion and Conclusions}

We have presented an evidence for correlated X-ray emission with the diffuse VSSRS 
in Abell~85. What are the possible emission processes (other than the IC/3K) that can give rise to
the excess X-ray emission? The
possibility that the emission is either the extension of the 
synchrotron radiation to the $\sim$1 keV X-ray band, or that arises from the synchrotron 
self Compton process (Rees 1967, Harris et al. 1994), can be ruled out from the observed 
very steep
spectral shape for $\nu \ge $ 1 GHz, signifying a dearth of necessary high energy photons.
Further possibilities that the emission is generated by an active galaxy or that it
arises from radio-jet cluster-medium 
interaction also appear remote due to the relaxed, `halo' type morphology of the 
VSSRS, and the absence of any optical counterpart.
 
Recent X-ray imaging and optical spectroscopy of clusters have shown the
presence of thermal X-ray emission associated with secondary substructures
possibly in the process of gravitational merger (e.g. Briel et al. 1991, Mohr et
al. 1993, Henry and
Briel 1993, Burns et al. 1994). Significantly, the
presence of radio-halo sources appears to be correlated with the evidence for
recent mergers, very high gas temperatures (7--14 keV), large velocity dispersions
($\approx 1300$~km~s$^{-1}$), and the absence of both cooling flows and a single dominant central galaxy (BM type
II or III; Edge et al. 1992, Tribble 1993, Feretti and Giovannini 1996). It has been proposed that these
data possibly indicate the heating of intracluster gas and the reaccelaration of cosmic ray
particles with stochastic amplification of magnetic fields, powered by the energy released in
the ongoing mergers (Tribble 1993). Can the excess X-ray emission seen with the VSSRS in A85
come from such a process?

Although imaging data alone can not resolve this question, it is interesting to note that 
in terms of its physical properties, unlike the 
other radio-halo host clusters such as the Coma, Abell~2163, 2218, 2255,2256, and 2319, the
cluster Abell~85 contains a central dominant cD galaxy, a central cooling flow of about 100
M$_{\odot}$~yr$^{-1}$, and relatively `cool' gas at a temperature of T $\sim$ 4 keV (Pislar
et al. 1997; Lima Neto et al. 1997). 

Durret et al. (1998) have detected a filamentary structure, visible both in
optical and in X-rays, linking Abell 85 to the neighbouring cluster Abell 87. They
present evidence for a possible merger of matter in this filament with the
southern region of Abell 85, in the `south-blob' region. Markevitch et al. (1998) detect an enhancement in the gas temperature in the same region, possibly
related to the merger. The position of the VSSRS, however, is not located there as it is shifted about
3 arcmin (300$h_{50}^{-1}$ kpc) to the northwest from the supposed shocked
region and has a cooler temperature for the thermal matter in its vicinity.

The resolution of the question of exact physical process behind the excess X-ray emission
would require  high spectral (and spatial) resolution imaging spectroscopy data. In the hard
X-ray energy regime of $\approx$ 10--20 keV, the contribution from thermal bremsstrahlung would
drop sharply, whereas the non-thermal inverse Compton flux would be visible as an extra component
with a steep power law spectrum of spectral index $\alpha \approx$ 1. Based on the data presented
in this work, we predict the possible IC/3K flux of $\approx 2.8 \times
10^{-13}$ erg s$^{-1}$ cm$^{-2}$, and the thermal bremsstrahlung flux of
$\approx 3.0 \times 10^{-13}$ erg s$^{-1}$ cm$^{-2}$ in
the hard X-ray band of 5--10 keV. These fluxes are well within the reach of the currently operative
BeppoSAX  telescope and the upcoming AXAF and the XMM missions. The nature of
the other excess X-ray and the steep
spectrum radio emission detected over the `south-blob' (cf. below) could
possibly also be understood with the spectroscopic X-ray data.

With the evidence currently available to us, energetically 
the most feasible mechanism for the X-ray emission from the VSSRS appears 
to be the IC/3K process. From 
theory (e.g. Ginzburg 1989), the (total) energy loss ratio can be shown to be ${f_{IC} 
\over f_{S}} \approx U_{rad}/ {B^{2} \over {8 \pi}}$, where $U_{rad} \approx 5 
\times 10^{-13}$ erg~cm$^{-3}$ is the energy density of the cosmological black-body radiation field, and 
${B^{2} \over {8 \pi }} \approx 4 \times 10^{-14}$ erg~cm$^{-3}$ is the observed 
magnetic energy density. Therefore, if the observed radiation is from IC/3K 
process, we expect $f_{IC} / f_{S} \approx 13$, which is comparable to the observed 
ratio $f_{IC} / f_{S} \approx 10$ (the difference is mainly attributed to the finite 
bandwidths of our data). If the excess X-rays are produced by the thermal bremsstrahlung process, it 
is difficult to understand why the observed ratio is comparable to the ratio expected if IC/3K were
the emission mechanism. This again
suggests that the detected X-rays are indeed a 
product of inverse Compton scattering. This data on ratio of fluxes has enabled us to obtain a
model independent magnetic field value, $B = 0.95 \pm 0.10 \mu$G, for the diffuse
emission volume, located at
$\sim 700 h_{50}^{-1}$ kpc from the cluster centre.

It is apparent (Fig. 2) that another diffuse X-ray excess (the `south-blob') is located
at $\alpha = 00^{\rm h}~41.8^{\rm m}$, $\delta = -09^{\rm d}~27.5^{\rm m}$. In the same
region, two radio components, without any definite optical counterparts, are observed with the
OSRT with $> 20 \sigma$ signal (162 mJy: northern source, 148 mJy: southern source). A spectral
index limit, $\alpha \ > 2.3$, is obtained for each of them, based on their non-detection in the
new VLA 1.4 GHz sky-survey (Condon et al. in preparation) to the $\approx$ 5 mJy flux density limit.  
The nature of
this excess X-ray emission is currently not well understood (Lima Neto et al. 1997). Is it
possible that a second IC/3K X-ray source is located here, co-spatial with yet another relic radio
plasma? If true, this can provide another magnetic field estimate at $\approx 1 h_{50}^{-1}$ Mpc from the
cluster centre. Detailed, low-frequency ($<327$ MHz) radio and X-ray spectral data 
are necessary to explore this possibility. 

Finally, the estimated magnetic field and the observed radiative fluxes from the VSSRS 
imply the presence of relativistic electrons with Lorentz factors $\gamma \approx$ 700--1700 
(for 0.5--2.4 keV band, $\nu_{x} \propto \nu_{bg} \gamma^{2} $). The population of
electrons in this energy range would in turn 
produce significant radio emission ($\sim $ 1400 Jy at 2 MHz) 
in the presently unexplored frequency range 
$\nu_{syn} \approx $ 2--10 MHz ($\nu_{syn} \propto B\;\gamma^{2}$). The radiative lifetime of such
`relic'
electrons is $\ga 10^{9}$ yr (Harris \& Grindlay 1979), comparable to the time-scale for
evolution of clusters. Although this frequency range is below the ionospheric
cutoff, the future generation radio telescopes observing from above the earth's
atmosphere or from the far side of the moon (Burns 1990), could make such
measurements. Such low frequency radio data in association with sensitive X-ray
data would prove invaluable in probing the physics of the intracluster media in
large scale structures. 

It is encouraging to note that the inverse Compton
X-ray technique has the great potential, not only in measuring the elusive intra-cluster magnetic fields,
but also in
probing the hitherto unexplored population of very old relativistic electrons  residing in the diffuse
intra-cluster space.

\section{Acknowledgments}
We thank F. Durret and D. Gerbal for fruitful comments and discussion.
G. Swarup, V.K. Kapahi, and late M.N. Joshi are greatfully thanked for
the OSRT radio data.


\begin{thebibliography}{}

\bibitem[\protect\citename{Bagchi} 1992]{Bagchi}
Bagchi J. 1992, Ph.D. Thesis, Dept. of Physics, Indian Institute of Science,
Bangalore, India
\bibitem[\protect\citename{Baldwin \& Scott} 1973]{Baldwin}
Baldwin J.E., Scott P.F., 1973, MNRAS, 165, 259
\bibitem[\protect\citename{Baylis et al.} 1967] {Baylis}
Baylis W.E., Schmidt W.M., Luscher E., 1967, Zeit. fur Astr. 66, 271
\bibitem[\protect\citename{Bazzano et al.} 1990] {Bazzano}
Bazzano A., Fusco-Femiano R., Ubertini P., Perrotti F., Quadrini E., Court 
A.J., Dean A.J., Dipper N.A., Lewis R., Stephen J.B., 1990, ApJ 362, L51  
\bibitem[\protect\citename{Blumenthal and Gould} 1970] {Blumenthal}
Blumenthal G.R., Gould R.J., 1970, Rev. Mod. Phys., 42(2), 237
\bibitem[\protect\citename{Briel et al.} 1991] {Briel10}
Briel U.G., Henry J.P., Schwarz R.A., Bohringer H., Ebeling H., Edge A.C.,
Hartner G.D., Schindler S., Trumper J., 
Voges W., 1991, A\&A 246, L10
\bibitem[\protect\citename{Burns} 1990] {Burns}
Burns J.O., 1990, Low Frequency Astrophysics from Space (Lecture Notes in
Physics 362, Springer-Verlag), 19
\bibitem[\protect\citename{Burns et al.} 1994] {Burns4}
Burns J.O., Rhee G., Owen F.N., Pinkney J., 1994, ApJ 423, 94 
\bibitem[\protect\citename{Dickey \& Lockman} 1990]{Dickey}
Dickey J.M., Lockman F.J., 1990, ARA\&A 28, 215
\bibitem[\protect\citename{Durret et al.} 1998] {Durret}
Durret F., Forman W., Gerbal D., Jones C., Vikhlinin A., 1998, A\&A in press
(astro-ph/9802183)
\bibitem[\protect\citename{Edge et al.} 1992] {Edge}
Edge A., Steward G.C., Fabian A.C., 1992, MNRAS 258, 177 
\bibitem[\protect\citename{Feenberg \& Primakoff} 1948] {Feenberg}
Feenberg E., Primakoff H., 1948, Phys. Rev. 73, 449
\bibitem[\protect\citename{Feigelson} 1995] {Feigelson}
Feigelson E.D., Laurent-Muehleisen S.A., Kollgaard R.I., Fomalont E.B., 1995,
ApJ 449, L149
\bibitem[\protect\citename{Felten \& Morrison} 1966] {Felten}
Felten J.E., Morrison P., 1966, ApJ 146, 686
\bibitem[\protect\citename{Feretti \& Giovannini} 1996] {Feretti}
Feretti L., Giovannini G., 1996, IAU Symp. 175
``Extragalactic Radio Sources'', p.~333 (Ekers R., et al. eds.)
\bibitem[\protect\citename{Forman \& Jones} 1984] {Forman}
Forman W., Jones C., 1982, ARA\&A 20, 547
\bibitem[\protect\citename{Ginzburg} 1989] {Ginzburg}
Ginzburg V.L., 1989, Application of Electrodynamics in Theoretical Physics
and Astrophysics, p.~431 (N.Y.: Gordon \& Breach)
\bibitem[\protect\citename{Goldshmidt \& Rephaeli} 1994] {Goldshmidt}
Goldshmidt O., Rephaeli Y., 1994, ApJ 431, 586  
\bibitem[\protect\citename{Harris \& Romanishin} 1974] {Harris1}
Harris D.E., Romanishin W., 1974, ApJ 188, 209
\bibitem[\protect\citename{Harris \& Grindlay} 1979] {Harris2}
Harris D.E., Grindlay J.E., 1979, MNRAS 188, 25
\bibitem[\protect\citename{Harris et al.} 1994] {Harris3}
Harris D.E., Carilli C.L., Perley R.A., 1994, Nature 367, 713
\bibitem[\protect\citename{Harris et al.} 1995] {Harris4}
Harris D.E., Willis A.G., Dewdney P.E., Batty J., 1995, MNRAS 273, 785
\bibitem[\protect\citename{Henry and Briel} 1993] {Henry}
Henry J.P., Briel U.G., 1993, Adv. Space Res. 13, (12)191 
\bibitem[\protect\citename{Hoyle} 1965] {Hoyle}
Hoyle F., 1965, Phys. Rev. Lett. 15, 131
\bibitem[\protect\citename{Joshi et al.} 1986] {Joshi}
Joshi M.N., Kapahi V.K., Bagchi J., 1986, Proc. of NRAO Workshop on Radio Continuum
Processes in Clusters of Galaxies, p.~73 (O'Dea C.P., Uson J.M eds.)
\bibitem[\protect\citename{Kaneda} 1995] {kaneda}
Kaneda H., Tashiro M., Ikebe Y., Ishisaki Y., Kubo H., Makishima K., Ohashi
T., Saito Y., Tabara H., Takahashi T., 1995, ApJ 453, L16
\bibitem[\protect\citename{Kim et al.} 1991] {Kim}
Kim K.-T., Tribble P.C., Kronberg P.P., 1991, ApJ 379, 80
\bibitem[\protect\citename{Kronberg} 1994] {Kronberg}
Kronberg P.P., 1994, Rep. Prog. Phys. 57, 325
\bibitem[\protect\citename{Laurent} 1994] {Laurent}
Laurent-Muehleisen S.A., Feigelson E.D., Kollgaard R.I., Fomalont E.B., 1994,
in: ``The Soft X-Ray Cosmos'',AIP Conference Proceedings 313,  p. 418,
(Schlegel E.M., Petre R. eds.)
\bibitem[\protect\citename{Lima Neto et al.} 1997]{LimaNeto}
Lima Neto G.B., Pislar V., Durret F., Gerbal D., Slezak E.,
1997, A\&A 327, 81
\bibitem[\protect\citename{Mather} 1994] {Mather}
Mather J.C. et al., 1994, ApJ 420, 439
\bibitem[\protect\citename{Markevitch et al.} 1998] {Markevitch}
Markevitch M., Forman W.R., Sarazin C.L., Vikhlinin A., 1998, ApJ Submitted (astro-ph/9711289)
\bibitem[\protect\citename{Mohr et al.} 1993] {Mohr3}
Mohr J.J., Fabricant D.G., Geller M.J., 1993, ApJ 413, 492
\bibitem[\protect\citename{Pacholkzyk} 1977] {Pacholkzyk}
Pacholkzyk A.G., 1970, Radio Astrophysics, p.~171 (WH Freeman \& Co., San Francisco)
\bibitem[\protect\citename{Pislar et al.} 1997]{Pislar}
Pislar V., Durret F., Gerbal D., Lima Neto G.B., Slezak E., 
1997, A\&A 322, 53
\bibitem[\protect\citename{Rees} 1967] {Rees}
Rees M.J., 1967, MNRAS 137, 429
\bibitem[\protect\citename{Rephaeli} 1977] {Rephaeli0}
Rephaeli Y., 1977, ApJ 212, 608
\bibitem[\protect\citename{Rephaeli et al.} 1987] {Rephaeli1}
Rephaeli, Y., Gruber D.E., Rothschild R.E., 1987, ApJ 320, 139
\bibitem[\protect\citename{Rephaeli \& Gruber} 1988] {Rephaeli2}
Rephaeli Y., Gruber D.E., 1988, ApJ 333, 133
\bibitem[\protect\citename{Rephaeli et al.} 1994] {Rephaeli}
Rephaeli Y., Ulmer M., Gruber D.E., 1994, ApJ 429, 554
\bibitem[\protect\citename{Rephaeli} 1995] {Rephaeli3}
Rephaeli Y., 1995, ARA\&A 33, 541
\bibitem[\protect\citename{Schwab} 1980] {Schwab}
Schwab F.R., 1980, Proc. SPIE 231, 18
\bibitem[\protect\citename{Slee \& Reynolds} 1984] {Slee}
Slee O.B., Reynolds J.E., 1984, Proc. Astr. Soc. Aust 5, 516
\bibitem[\protect\citename{Sukumar et al.} 1988] {Sukumar}
Sukumar S., Velusamy T., Rao P., Swarup G., Bagri D.S., Joshi M.N.,
Ananthkrishnan S., 1988,
Bull. Astron. Soc. India 16, 93
\bibitem[\protect\citename{Swarup} 1984]{Swarup}
Swarup G. 1984, JA\&A 5, 139
\bibitem[\protect\citename{Tribble} 1993] {Tribble}
Tribble P.C., 1993, MNRAS 263, 31
\bibitem[\protect\citename{Zimmermann et al.} 1994]{Zimmermann}
Zimmermann H.U., Becker W., Belloni T., D\"obereiner S., Izzo C.,
Kahabka P., Schwentker O., 1994, EXSAS 
Users' Guide, MPE Report 244
\label{lastpage}
\end{thebibliography}
\end{document}